\RequirePackage{fix-cm}
\documentclass[smallextended]{svjour3}       
\smartqed  
\usepackage{graphicx}
\usepackage{epsfig}
\usepackage{hyperref}
\usepackage{color}
\usepackage{amsmath}
\newcommand{\be}{\begin{equation}}
\newcommand{\ee}{\end{equation}}

\newcommand{\eV}{\;\mathrm{eV}}
\journalname{JOSC}
\begin{document}

\title{High-temperature superconductors as ionic metals
\thanks{Supported by the Croatian Science Foundation under Project IP-2018-01-7828.}
}


\author{D.~K.~Sunko
}


\institute{D.~K.~Sunko \at
Department of Physics, Faculty of Science, University of
Zagreb,\\ Bijeni\v cka cesta 32, HR-10000 Zagreb, Croatia.\\
              \email{dks@phy.hr}           
}

\date{Received: date / Accepted: date}

\maketitle

\begin{abstract}

High-temperature superconductors are reviewed in light of the fact that their binding energy is ionic. The conducting electrons are dominated by the much larger energy scales coming from ligand Coulomb integrals, including the out-of-plane ones, which are responsible for the Fermi arcs. The historic reinterpretation of Hund's rule from an intraelectronic to a central mean-field effect is applied to compare the cuprates to the pnictides. It is argued that the cuprates conform to the now-standard central-field paradigm, while the generally abandoned intraelectronic mechanism is exceptionally applicable to the pnictides. A non-adiabatic Fermi liquid paradigm is inferred from the phenomenological evidence. Glueless superconductivity is interpreted as the limiting case of Cooper-pair scattering in cuprates when the Cu ion is perfectly rigid.

\end{abstract}

\section{Introduction}

No generally accepted explanation exists for the superconductivity (SC) of cuprate perovskites~\cite{Bednorz86} and ferropnictides~\cite{Kamihara06}. The exceptionally large effort involved, both experimentally and theoretically, without a resolution over thirty years, naturally raises questions about the effort itself. Several recent synthetic works~\cite{Mazumdar18,Drozdov18,Pelc19} indicate that mainstream theoretical approaches may have adopted a reductionist attitude too soon. In particular, the much-emphasized strong electronic correlations in the cuprates were dominantly studied in the minimal two-dimensional, one-band model of intraelectronic copper-site repulsion~\cite{Anderson90}.

There is an alternative approach, also as old as cuprate research itself, of enriching the context, in the hope of reaching a complete physical narrative before reduction. One considers the actual chemical composition of the high-temperature superconductors in an attempt to establish the main features of the physical regime required for high-T$_c$ superconductivity to occur~\cite{Mazumdar89,Barisic90,Eskes93}. At the interface of physics and chemistry, there appears an exciting new scientific frontier, drawing on insights from both fields, to which the usual term ``materials science'' does not do justice if it is only associated with the operational level of density-functional calculations (DFT). The present work is a personal contribution to this program on the occasion of Theodore Henry Geballe's 100th birthday. Established results and new insights are presented in a spirit of wide eclecticism, familiar to his many admirers. In the vast body of his work, I mention two articles below, separated by fifty years, which impinge directly on questions raised in the present essay~\cite{Geballe63,Marezio13}.

\subsection{Ionic metals}

Metallic binding means that the only way to provoke a metal-insulator transition is to evaporate the sample. Mott~\cite{Mott49} was the first to consider the metal-insulator transition as a function of \emph{inter}-atomic distance in model lattices, comparing Coulomb repulsion in the tails of electron wave functions, centered at different atoms, with the energy gain of delocalization. The inter-atomic (ligand!) Coulomb integral drives the metal-insulator transition as the conducting liquid vaporises.

The term Mott transition usually refers to a different mechanism~\cite{Boer37}, by which the \emph{intra}-atomic repulsion in the metal $d$-orbital of transition-metal oxides makes them insulating at half-filling. However, in the latter case, the binding is ionic. The metallic contribution to the formation energy is negligible to zeroth order whenever metal-insulator transitions (with doping, or pressure) occur without significant structural changes. Conducting materials in this regime may be called \emph{ionic metals}, a limiting case of ionocovalency. Both superconducting cuprates and pnictides are in this limit.

\subsection{Hund's rule}

Between the 1930's and the 1960's, it was generally believed that the ubiquitous Hund's rule had been understood along the lines originally proposed by Slater~\cite{Slater29}: the space part of the two-electron wave function lowers Coulomb repulsion in the triplet state. It was then discovered that this explanation was wrong~\cite{Davidson65}. The spin-symmetric states are preferred because the fully antisymmetric space wave function screens the nucleus least efficiently, so that all electrons can benefit from a contraction of the atomic radius. The real driving mechanism is electron-nucleus attraction, not electron-electron repulsion~\cite{Yamanaka05}. Because of the nuclear charge factor $Z$, the former effect tends to be much larger than the latter, which misleadingly gives the same conclusion. Even for the triplet state of helium, Hund's rule was obtained with $s$-wave atomic functions alone, simply by allowing the one-body wave functions in the singlet and triplet states to have different Bohr radii~\cite{Sajeev08}. This result runs contrary to the plausible idea that the two electrons need to stay on opposite sides of the nucleus, i.e.\ be angularly correlated, in order to minimize mutual repulsion. In the present context, such a mechanism may be called ``glueless''~\cite{Anderson07}, because it achieves the same outcome by average (radial) energy balance as would be achieved by a much weaker scattering (angular) mechanism.

Based on the above insight, one may distinguish two kinds of strong electronic correlations in ionic metals. One is radial or mean-field, encoded e.g.\ by the +U correction in DFT calculations. The other is angular, or strongly correlated in the strict sense, typically described by dynamic mean-field theory (DMFT). It will be argued below that the first is dominant in the cuprates, the second in the pnictides.

\subsection{Fermi liquid}

Landau~\cite{Landau56} obtained the equation of state for liquid $^3$He in the hydrodynamic limit,
\be
n(\mathbf{r},\mathbf{p})=\frac{1}{e^{(\varepsilon(\mathbf{r},\mathbf{p})-\mu)/kT}+1},
\label{fl}
\ee
by adding a fermionic correction $(1-n)\ln (1-n)$ to Boltzmann's H-theorem. It refers to thermodynamic quantities~\cite{Heinonen87}: $n(\mathbf{r},\mathbf{p})$ is the density of particles of momentum $\mathbf{p}$ in a macroscopic small volume of liquid centered at $\mathbf{r}$, while  $\varepsilon(\mathbf{r},\mathbf{p})=\delta E/\delta n$ is the variation of the internal energy with this density~\cite{Leggett06}.

Eq.~\eqref{fl} can be derived~\cite{Nozieres62-1} under the assumption that many-body perturbation theory is valid to all orders, which is a formal expression of Landau's adiabatic argument. However, like all thermodynamic equations of state, it refers to observations directly. Whenever experiments show a Fermi surface, some density in the sample satisfies Eq.~\eqref{fl}, whether it can be derived adiabatically or not. It will be pointed out below that the Fermi arcs~\cite{Yoshida12} have in fact been obtained in a one-body microscopic model, once the out-of-plane Coulomb integrals were properly taken into account~\cite{Lazic15}.

\subsection{Electronic structure}

The full 3D structure of cuprates was taken into account in a comprehensive study~\cite{Pavarini01} which established the hierarchy of reduction for tight-binding models describing the functional copper-oxide planes. The smallest three-dimensional, chemically realistic model has six bands. Three come from the planar Cu $3d_{x^2-y^2}$ and O $2p_x$ and $2p_y$ orbitals. Three are due to the Cu $4s$ and $d_{z^2}$ orbitals and the apical O $2p_z$ orbital, which connect the planar electrons to the third dimension. The latter can be fused into an ``effective $4s$ orbital'' to obtain a still-realistic four-band 2D model.

The first departure from chemical realism is to drop the $4s$ orbital, giving the well-known three-band Emery model~\cite{Emery87}. Physically, it conflates two hopping mechanisms between the planar oxygens. One is the direct $2p_x$--$2p_y$ overlap $t_{pp}$, of the order of $0.1\eV$. The other is the second-order process $2p_x$--$4s$--$2p_y$, which benefits from the very large overlap $t_{ps}\sim 2\eV$, despite the $4s$ orbital being $\Delta_{ps}\sim 4$--$6\eV$ away from the $2p_{x,y}$ (and Fermi) levels. It can be estimated as $t_{ps}^2/\Delta_{ps}\sim 0.7$--$1\eV$, which explains why the effective value of $t_{pp}$, needed e.g.\ to fit ARPES in the three-band model~\cite{Sunko07}, appears unrealistically large. The three-band model departs from chemical reality quantitatively, but not qualitatively.

One can reduce further to a one-band model with more than nearest-neighbor hoppings, the so-called $t$--$t'$ or $t$--$t'$--$t''$ model. Here the $t'$ term is required for particle-hole symmetry breaking, due to the O orbitals in reality. It must change sign between the hole- and electron-doped cuprates in order to fit ARPES~\cite{Kyung04}, indicating a qualitative departure from the underlying chemistry, because a structural overlap cannot change sign with doping. It will be pointed out below that, although it is nominally  two-dimensional, the four-band model still allows one to understand a crucial dimensional crossover, occuring at the Fermi surface of all cuprate superconductors.

\section{Strong correlations in high-T$_c$ superconductors}

\subsection{Fermi arcs}

Fermi arcs in cuprates have been the object of much scrutiny~\cite{Yoshida12}, in particular from the point of view of strong correlations. In a weak-coupling approach, the Fermi surface either closes on itself, or touches the edge of the zone, so it must be a strong-correlation effect when it does neither. It then appears natural to claim that the electrons in the arc are not a Fermi liquid. Surprisingly, the first conclusion holds, but the second does not, as shown by counterexample, specifically in La$_{2-x}$Sr$_x$CuO$_4$ (LSCO). It is possible to construct a Fermi liquid whose Fermi surface consists of arcs~\cite{Lazic15}.

The key to such a construction is to model the out-of-plane dopands as real physical atoms, which is quite demanding computationally because it involves large unit cells in a DFT calculation. Such a large cell corresponds to a small Brillouin zone in the plane. In this zone, there are numerous Fermi surfaces which are all closed or touch the edge, because DFT is a one-body approach. In order to map the results onto the usual large zone of the CuO$_2$ planes, a standard band-unfolding algorithm is used, originally developed for disordered alloys~\cite{Popescu12}. It gives Fermi arcs just as observed, centered around the zone diagonal and growing with doping.

The logic of a counterexample turns any criticism of DFT on its head here: the more physically limited the calculation, the more difficult to escape the conclusion that the carriers in the real arcs are a Fermi liquid. The Brillouin zone is a kinematic construction, while the band-unfolding is just a projection, so the only \emph{physical} requirement for the calculation to be realistic is for the Coulomb field of the dopands to reach the planes, i.e.\ be unscreened between them, because that is how the dopand (large) unit cell becomes relevant for the in-plane carriers. The calculated wave functions verify this requirement. One finds, in laboratory space, a local polarization response in the atomic charge densities, stopping abruptly at both nearest CuO$_2$ planes, which screen the dopands efficiently in the perpendicular direction. Notably, in wave-function space, the Cu component of the metallic electrons on the arc consists of planar $d_{x^2-y^2}$ orbitals only, while the antinodal gapped state contains an essentially random superposition of various planar and perpendicular orbitals.

The physical picture emerging from this calculation is that the arcs are a \emph{simple kinematic effect}. The antinodal gap is not due to 2D intraelectronic correlations, but rather the opposite, the Fermi liquid in the arcs appears where the 3D charge-transfer insulator (``rock salt''), which would normally be expected to gap the whole zone, cannot make itself felt. Experimentally, it is possible to manipulate arc length by out-of-plane dopands at fixed carrier concentration~\cite{Okada08}.

This interpretation of ``arc protection''~\cite{Ferrero09} is verified analytically~\cite{Lazic15} within the four-band model mentioned above. The secular polynomial of this model factorizes along the zone diagonal ($k_x=k_y$), so that the dispersion along the zone diagonal is uncoupled from the effective Cu $4s$ orbital which is the physical conduit of out-of-plane effects on the plane. In other words, the zone diagonal is orthogonal to the third dimension \emph{in wave-function space}. Hence Coulombic 3D effects which otherwise determine the bulk properties cannot affect it. Towards the edge of the zone, the 3D character of the material is manifested in the four-band model as an increasing coupling to the $4s$ orbital. In the full DFT calculation, it is accompanied with increasing orbital disorder.

The small Brillouin zone pertaining to the large dopand unit cell is not observed because of dopand disorder, which was not included in the calculation~\cite{Lazic15}. This is just the disordered-alloy situation for which the band-unfolding algorithm~\cite{Popescu12} was originally developed. The zone of a parent metal persists spectroscopically long after alloying has nominally destroyed translational invariance. The disorder effects appear modestly, as a gradual evolution of the peaks in the robust initial zone set by the parent~\cite{Popescu12}. In the cuprate samples, the large zone is similarly expected to average out the disordered small ones, remaining the only one to be observed spectroscopically.

\subsection{DFT vs.\ DMFT}

In the above calculation, the large on-site repulsion $U$ within the copper $d$ orbital was treated with the usual +U correction
\be
U\,\mathrm{tr}_d(\rho -\rho^2)
\label{plusU}
\ee
where the trace is over the $d$ orbitals only. Because the total density matrix is idempotent, $\rho^2=\rho$, the trace is identically zero if extended over all orbitals in the calculation. As it stands, the correction penalizes departures from the idempotent occupation numbers, zero and one, in the $d$ orbitals only, thus pushing the Cu orbitals towards the ionic limit. With a value $U=4\eV$, known to give the correct formation energy of LSCO, the measured optical gap of $1.8\eV$ was obtained at half-filling, a useful reality check on the calculation~\cite{Lazic15}.

In the context of Hund's rule, the success of the +U correction means that the Cu orbital has compensated the strong intraatomic correlation radially, by transferring charge to the ligand oxygen, which is itself not strongly correlated. The crucial requirement for this effect is structural, namely that the O sites bridge the Cu sites, as they do in the cuprates, precluding a direct Cu--Cu hybridization. This structure is reflected in a $2p$--$3d$ overlap, which in turn allows the correction~\eqref{plusU} to be non-zero, because the sum over $d$ orbitals does not include the oxygen $p$ ones, required to establish the idempotence of the density matrix. Notably, the same Cu--O bond determines the size of the unit cell in the plane, so the standard Hund's rule driving mechanism, namely the size of the atom, can be said to be active here as well, \emph{mutatis mutandis}.

Compare this to the pnictides, specifically (Ba,K)Fe$_2$As$_2$ (KFeAs)~\cite{Fink09}. DFT calculations show that the As ligands and the Fe $e_{2g}$ orbitals which hybridize with them are strongly ionic~\cite{Eschrig09}. The corresponding bands are at high energy, i.e.\ deep below the Fermi surface. On the other hand, the bands crossing the Fermi surface are purely Fe $t_{2g}$ in character. They are due to direct hopping between the Fe atoms, caged in the As tetrahedra. The DFT reflection of that situation is, apropriately, that the +U correction becomes ineffective. The $t_{2g}$ orbitals do not hybridize with any others, so that the density matrix must be idempotent in their subspace alone, while the $e_{2g}$'s are already ionic, driving the correction~\eqref{plusU} to zero. Structurally, this means that the size of the unit cell is determined by the high-energy ligands, so that the metallic Fe orbitals find themselves as if in an \emph{atom of fixed size}.

With radial relaxation turned off, the strong correlation effects in the $t_{2g}$ bands are driven solely by intraelectronic repulsion. They are expected to be angular, as in the original explanation of Hund's rule. Consistently with that interpretation, a $0.2\eV$ discrepancy with experiment in the position of a $t_{2g}$ band maximum~\cite{Borisenko16} could only be resolved with a DMFT calculation~\cite{Derondeau17}, which brought it to the Fermi surface as observed. Effectively, $t_{2g}$ orbitals are in a one-band Hubbard model situation, with neighboring sites strongly correlated, and lattice effects ignored.

Notably, the measured effect of strong correlations is nearly ten times smaller than in LSCO, where the optical gap of $1.8\eV$, and the arcs, are accounted for by the +U correction~\cite{Lazic15}. This difference is again consistent with the chemical narrative, where radial effects are systematically larger than angular ones. The latter dominate in KFeAs only because one set of orbitals fixes the size of the unit cell ionically, while another, structurally ``slaved'' to it, is metallic. In cuprates, the same orbitals are responsible for both, so the radial effects dominate the angular ones as usual.

\subsection{Sulphur hydride}

There is little doubt at present, that the mechanism of SC in sulphur hydride (H$_3$S)~\cite{Drozdov15} is phononic~\cite{Gorkov16}, so it would appear that the above discussion has little to do with it. Nevertheless, an intriguing parallel presents itself. H$_3$S has two high-T$_c$ phases, a 100~K phase below 150~GPa, and a 200~K phase above. The structural transition between the two~\cite{Duan14} is remarkable from the present point of view.

In the 200~K phase, the H atoms bridge positions of S atoms in a cubic lattice. It is obvious from this structure that pressure affects the H--S charge transfer through the corresponding interatomic overlap, responsible for the lattice constant. This ``radial'' situation is directly analogous to the cuprates.

On the other hand, the 100~K phase consists of individual H$_3$S tetrahedra, themselves arranged in a cubic lattice. Again, it is clear from the structure that pressure will affect the distance between these tetrahedra before it changes the the H--S distance within them. The S atom is in the fixed-size, or ``angular'', situation now, like in the pnictides --- and T$_c$ drops by half.

\section{Discussion}

In light of the program outlined in the introduction, one should ask, does materials science bring us any closer to understanding high-T$_c$ superconductivity. There are several questions involved. Is the high-T$_c$ mechanism in the cuprates and the pnictides the same, or different? Are the strong correlation effects essential to it, or just a distraction? Is the BCS mechanism relevant, or not?

A basic property of the BCS mechanism is that it can use \emph{any} scattering in the particle-particle channel as long as it affects free carriers on the Fermi surface, i.e.\ is not gapped. If there are several such mechanisms, they add up to raise T$_c$. The physical reason for this lack of discrimination is that Cooper pairs are not bound, so the carriers travel as free particles between scattering events~\cite{Mahan90}, able to benefit from any mix of interactions. Thus it is not physically natural to restrict the BCS mechanism terminologically to phonon-mediated interactions. It is relevant whenever the SC instability is by Cooper pairing, whatever the origin of the free-carrier scattering. For example, the glueless scenario~\cite{Anderson07} is not alternative to BCS, because it invokes a condensate of Cooper pairs. Alternative scenarios have an alternative picture of the condensate.

Models with bound pairs are alternatives to BCS, in which the condensate is a superfluid of such pairs. They have mostly been based on a polaronic mechanism~\cite{Alexandrov96}. However, the polaronic response is not engaged if the hopping is fast~\cite{Friedel96}. A fast O--O hopping channel was identified~\cite{Pavarini01} in cuprate ionic metals, which makes the effective overlap $t_{pp}$ large. Furthermore, the carriers on the zone diagonal do not see the $c$-direction at all~\cite{Lazic15}. The strongly ionic background determines the properties of cuprate carriers in such a way that they are not slowed down by the lattice. Given that the wide O band also implies a high Fermi energy, the materials-science insights are in favor of SC by Cooper pairing in the cuprates.

The other two questions can be coupled. Suppose one were told that the SC mechanism in cuprates and pnictides is the same. Would one be forced to conclude that the strong correlations were a distraction? The materials-science analysis allows one, somewhat surprisingly, to answer in the negative. One can imagine instead, whatever the Coulomb correlations are doing well in the radial channel, they are doing half as well in the angular one. This conjecture is in accord with the history of Hund's rule, where the radial mechanism was hidden in plain sight for forty years, simply because the weaker angular mechanism provided the same outcome. It brings into focus a distinction between two issues concerning the SC mechanism. One is the reason for T$_c>0$, i.e.\ that SC appears at all, the other the reason why T$_c$ is high.

The first issue centers on the carrier states involved. For example, alkali metals are Fermi liquids which do not superconduct. Strikingly, they become superconducting under moderate pressure just as their band structure departs from the free-electron picture, by acquiring a significant $p$ or $d$ component, due to overlap of ionic cores~\cite{Schilling06}. Studies of SC in noble metals and their alloys~\cite{Geballe63,Raub84} offer extensive evidence that the orbital state of the carriers affects SC qualitatively. In the cuprates, it has long been established experimentally that the oxygen orbitals must be opened for SC to occur, both on the hole-doped~\cite{Nucker87,Tranquada87,Fujimori87} and electron-doped~\cite{Barlingay90} sides of the phase diagram. In addition, there may be no site-energy splitting between the O $2p_x$ and $2p_y$ orbitals, whether by crystal field~\cite{Barisic90} or by impurities~\cite{Pelc15}.

The second issue centers on the SC scattering mechanism. The traditional reason to connect strong correlations with high T$_c$ is that they are obvious candidates to provide a high-energy scale for the SC scattering.

The comparison of cuprates and pnictides suggests that one needs to take into account both structural and dynamic roles of the large Coulomb integrals in the search for a high T$_c$. In both classes of materials, the structure ensures that the metallic carriers are not slowed down by polaronic effects. Specifically, in the cuprates the O--O hopping is mostly by a virtual high-energy process~\cite{Pavarini01}. In the pnictides, the metallic Fe--Fe hopping is by the $t_{2g}$ orbitals, which are of a different symmetry than those responsible for setting the lattice constant~\cite{Eschrig09}. Their common ground is a kinematic separation of a Fermi liquid from the low-energy background in the first step, which is then subject to a high-energy BCS scattering instability, raising issues of dynamics in the second step.

Concerning dynamics, the cuprates and pnictides are different, but not essentially so. In the cuprates, a recent systematic study~\cite{Pelc19} has proposed that the principal SC scattering is of O $2p_{x,y}$ holes on the Cu $3d^9$ state. This scattering channel has been called radial here. Such a scenario parallels the modern interpretation of Hund's rule, with Cu playing the nucleus to the O electron cloud. It becomes glueless in the limiting case of zero deformability of the Cu ion, corresponding to a purely electronic radial energy-balance calculation. Otherwise, the question is which internal states of Cu become important. (At least one measurement unexpectedly claims that these are the Cu $t_{2g}$ orbitals~\cite{Little07}.) In a strongly ionic material, such issues conflate the polaronic, excitonic, and chemical viewpoints at the level of a single unit cell, because different excitation and oxidation states of the ion can affect the local interatomic distance~\cite{Marezio13,Bianconi87}. A significant result in this vein is that the $d$-wave SC channel is not affected by the O$_x$--O$_y$ repulsion $V_{pp}$ because of different orbital symmetry~\cite{Valkov16}, testifying to the importance of chemical structure for the study of electron dynamics in ionic metals.

In the pnictides, the dominant interactions are intraelectronic, of the kind called angular here. The ensuing elaborate correlations give rise to effective magnetic terms up to next-nearest neighbors~\cite{Si16}. The much lower Coulomb scales are accompanied by a more familiar SC phenomenology, presumably because the physical Pippard scale $\sim v_F/\Delta$ (pair size) is several lattice spacings, so it can still be governed by self-consistent equations, while in cuprates it is cut off from below by the structural scale of the unit cell~\cite{Friedel96}.

It is encouraging that the ten times lower scattering scales in the pnictides translate into a SC T$_c$ only 2-3 times lower than in cuprates. This indicates that even a lower-energy scattering mechanism may be enhanced by a propitious chemical structure, which enforces the Fermi-liquid properties kinematically, decoupling them from the low-energy feedbacks relevant for normal metals. Indeed, when scaled to the number of carriers, resistivity in the cuprates follows a pure $T^2$ law over a much wider temperature range~\cite{Pelc19} than in textbook Fermi liquids such as In or Al, where the $T^2$ contribution must in addition be disentangled from a Gr\"uneisen crossover curve~\cite{Garland69}. In the pnictides, a $\sim 100$-K wide almost pure $T^2$ region is similarly found~\cite{Mukherjee15} at low temperatures.

Despite all their differences, cuprates and pnictides seem to lie in a similar ``sweet spot'' for SC: a Fermi liquid, uncoupled from the lattice, scattering by a high-energy mechanism which does not slow down the free carriers. It remains to be seen whether H$_3$S can be brought into the same fold. The question hinges on whether the strong electron-phonon coupling avoids polaronic effects~\cite{Gorkov16}, or not~\cite{Bussmann-Holder16}.

\section{Conclusion}

The natural framework for the SC mechanism in high-temperature superconductors is the BCS mechanism in its wider sense, with Cooper pairing but not necessarily via phonon interactions. Under the usual adiabatic Fermi-liquid construction, the SC scattering mechanism feeds back into the effective mass, leading to instabilities if it is too strong. An ideal scenario for high-temperature SC emerges from the comparative study of ionic metals, in which this feedback does not happen. Observations indicate a Fermi liquid interacting via a high-energy scattering mechanism, while its own properties are determined kinematically by the much-higher-energy ionic background. In the cuprates, the unscreened dopands are a critical part of that background. This scenario extends the Fermi-liquid concept into the non-adiabatic regime, accounting in particular for the Fermi arcs. In the SC state, a difference between cuprates and pnictides is that in the cuprates, not only the Fermi-liquid, but also the SC scales are affected by the structural ones, in particular the size of the unit cell.

\begin{acknowledgements}

Conversations with N. Bari\v{s}i\'c, I. Bo\v{z}ovi\'c, J. Fink, J.~E. Hirsch, Wei Ku, P. Lazi\'c, S. Mazumdar, D. Pavuna, E. Tuti\v{s} and T. Valla are gratefully acknowledged.

\end{acknowledgements}


\begin{thebibliography}{10}
\providecommand{\url}[1]{{#1}}
\providecommand{\urlprefix}{URL }
\expandafter\ifx\csname urlstyle\endcsname\relax
  \providecommand{\doi}[1]{DOI \discretionary{}{}{}#1}\else
  \providecommand{\doi}{DOI \discretionary{}{}{}\begingroup
  \urlstyle{rm}\Url}\fi

\bibitem{Bednorz86}
J.~Bednorz, K.~M\"uller, Z. Physik B - Condensed Matter \textbf{64}(2), 189
  (1986).
\newblock \doi{10.1007/BF01303701}

\bibitem{Kamihara06}
Y.~Kamihara, H.~Hiramatsu, M.~Hirano, R.~Kawamura, H.~Yanagi, T.~Kamiya,
  H.~Hosono, Journal of the American Chemical Society \textbf{128}(31), 10012
  (2006).
\newblock \doi{10.1021/ja063355c}

\bibitem{Mazumdar18}
S.~Mazumdar, Phys. Rev. B \textbf{98}, 205153 (2018).
\newblock \doi{10.1103/PhysRevB.98.205153}

\bibitem{Drozdov18}
I.K. Drozdov, I.~Pletikosi\'c, C.K. Kim, K.~Fujita, G.D. Gu, J.C.S. Davis, P.D.
  Johnson, I.~Bo\v{z}ovi\'c, T.~Valla, Nature Communications \textbf{9}(1),
  5210 (2018).
\newblock \doi{10.1038/s41467-018-07686-w}

\bibitem{Pelc19}
D.~Pelc, P.~Pop{\v c}evi{\'c}, M.~Po{\v z}ek, M.~Greven, N.~Bari{\v s}i{\'c},
  Science Advances \textbf{5}(1) (2019).
\newblock \doi{10.1126/sciadv.aau4538}

\bibitem{Anderson90}
P.W. Anderson, Phys. Rev. Lett. \textbf{64}, 1839 (1990).
\newblock \doi{10.1103/PhysRevLett.64.1839}

\bibitem{Mazumdar89}
S.~Mazumdar, in \emph{Interacting electrons in reduced dimensions}, ed. by
  D.~Baeriswyl, D.K. Campbell (Plenum Press, New York, 1989), pp. 315--329

\bibitem{Barisic90}
S.~Bari\v{s}i\'{c}, J.~Zelenko, Solid State Communications \textbf{74}(5), 367
  (1990).
\newblock \doi{10.1016/0038-1098(90)90504-5}

\bibitem{Eskes93}
H.~Eskes, J.H. Jefferson, Phys. Rev. B \textbf{48}, 9788 (1993).
\newblock \doi{10.1103/PhysRevB.48.9788}

\bibitem{Geballe63}
T.H. Geballe, B.T. Matthias, V.B. Compton, E.~Corenzwit, G.W. Hull, Phys. Rev.
  \textbf{129}, 182 (1963).
\newblock \doi{10.1103/PhysRev.129.182}

\bibitem{Marezio13}
M.~Marezio, O.~Chmaissem, C.~Bougerol, M.~Karppinen, H.~Yamauchi, T.H. Geballe,
  APL Materials \textbf{1}(2), 021103 (2013).
\newblock \doi{10.1063/1.4817895}

\bibitem{Mott49}
N.F. Mott, Proc. Phys. Soc. A \textbf{62}, 416 (1949)

\bibitem{Boer37}
J.H. de~Boer, E.J.W. Verwey, Proceedings of the Physical Society
  \textbf{49}(4S), 59 (1937).
\newblock \doi{10.1088/0959-5309/49/4s/307}

\bibitem{Slater29}
J.C. Slater, Phys. Rev. \textbf{34}, 1293 (1929).
\newblock \doi{10.1103/PhysRev.34.1293}

\bibitem{Davidson65}
E.R. Davidson, The Journal of Chemical Physics \textbf{42}(12), 4199 (1965).
\newblock \doi{10.1063/1.1695919}

\bibitem{Yamanaka05}
S.~Yamanaka, K.~Koizumi, Y.~Kitagawa, T.~Kawakami, M.~Okumura, K.~Yamaguchi,
  International Journal of Quantum Chemistry \textbf{105}(6), 687 (2005).
\newblock \doi{10.1002/qua.20784}

\bibitem{Sajeev08}
Y.~Sajeev, M.~Sindelka, N.~Moiseyev, The Journal of Chemical Physics
  \textbf{128}(6), 061101 (2008).
\newblock \doi{10.1063/1.2837456}

\bibitem{Anderson07}
P.W. Anderson, Science \textbf{316}(5832), 1705 (2007).
\newblock \doi{10.1126/science.1140970}

\bibitem{Landau56}
L.D. Landau, Sov. Phys. JETP \textbf{3}, 920 (1956)

\bibitem{Heinonen87}
O.~Heinonen, W.~Kohn, Phys. Rev. B \textbf{36}, 3565 (1987).
\newblock \doi{10.1103/PhysRevB.36.3565}

\bibitem{Leggett06}
A.J. Leggett, \emph{Quantum Liquids} (Oxford University Press, Oxford, 2006)

\bibitem{Nozieres62-1}
J.M. Luttinger, P.~Nozi\`eres, Phys. Rev. \textbf{127}, 1431 (1962).
\newblock \doi{10.1103/PhysRev.127.1431}

\bibitem{Yoshida12}
T.~Yoshida, M.~Hashimoto, I.~M.~Vishik, Z.X. Shen, A.~Fujimori, Journal of the
  Physical Society of Japan \textbf{81}(1), 011006 (2012).
\newblock \doi{10.1143/JPSJ.81.011006}

\bibitem{Lazic15}
P.~{Lazi\'c}, D.K. Sunko, EPL (Europhysics Letters) \textbf{112}(3), 37011
  (2015).
\newblock \doi{10.1209/0295-5075/112/37011}

\bibitem{Pavarini01}
E.~Pavarini, I.~Dasgupta, T.~Saha-Dasgupta, O.~Jepsen, O.K. Andersen, Phys.
  Rev. Lett. \textbf{87}, 047003 (2001).
\newblock \doi{10.1103/PhysRevLett.87.047003}

\bibitem{Emery87}
V.J. Emery, Phys. Rev. Lett. \textbf{58}, 2794 (1987)

\bibitem{Sunko07}
D.K. Sunko, S.~Bari\ifmmode \check{s}\else \v{s}\fi{}i\ifmmode~\acute{c}\else
  \'{c}\fi{}, Phys. Rev. B \textbf{75}, 060506(R) (2007).
\newblock \doi{10.1103/PhysRevB.75.060506}

\bibitem{Kyung04}
B.~Kyung, V.~Hankevych, A.M. Dar\'{e}, A.M.S. Tremblay, Phys. Rev. Lett.
  \textbf{93}, 147004 (2004)

\bibitem{Popescu12}
V.~Popescu, A.~Zunger, Phys. Rev. B \textbf{85}, 085201 (2012).
\newblock \doi{10.1103/PhysRevB.85.085201}

\bibitem{Okada08}
Y.~Okada, T.~Takeuchi, T.~Baba, S.~Shin, H.~Ikuta, Journal of the Physical
  Society of Japan \textbf{77}(7), 074714 (2008).
\newblock \doi{10.1143/JPSJ.77.074714}

\bibitem{Ferrero09}
M.~Ferrero, P.S. Cornaglia, L.~De~Leo, O.~Parcollet, G.~Kotliar, A.~Georges,
  Phys. Rev. B \textbf{80}, 064501 (2009).
\newblock \doi{10.1103/PhysRevB.80.064501}

\bibitem{Fink09}
J.~Fink, S.~Thirupathaiah, R.~Ovsyannikov, H.A. D\"urr, R.~Follath, Y.~Huang,
  S.~de~Jong, M.S. Golden, Y.Z. Zhang, H.O. Jeschke, R.~Valent\'{\i},
  C.~Felser, S.~Dastjani~Farahani, M.~Rotter, D.~Johrendt, Phys. Rev. B
  \textbf{79}, 155118 (2009).
\newblock \doi{10.1103/PhysRevB.79.155118}

\bibitem{Eschrig09}
H.~Eschrig, K.~Koepernik, Phys. Rev. B \textbf{80}, 104503 (2009).
\newblock \doi{10.1103/PhysRevB.80.104503}

\bibitem{Borisenko16}
S.V. Borisenko, D.V. Evtushinsky, Z.H. Liu, I.~Morozov, R.~Kappenberger,
  S.~Wurmehl, B.~Buchner, A.N. Yaresko, T.K. Kim, M.~Hoesch, T.~Wolf, N.D.
  Zhigadlo, Nat Phys \textbf{12}(4), 311 (2016).
\newblock \doi{10.1038/nphys3594}.
\newblock Letter

\bibitem{Derondeau17}
G.~Derondeau, F.~Bisti, M.~Kobayashi, J.~Braun, H.~Ebert, V.A. Rogalev, M.~Shi,
  T.~Schmitt, J.~Ma, H.~Ding, V.N. Strocov, J.~Min\'ar, Scientific Reports
  \textbf{7}(1), 8787 (2017).
\newblock \doi{10.1038/s41598-017-09480-y}

\bibitem{Drozdov15}
A.P. Drozdov, M.I. Eremets, I.A. Troyan, V.~Ksenofontov, S.I. Shylin, Nature
  \textbf{525}(7567), 73 (2015).
\newblock \doi{10.1038/nature14964}.
\newblock Letter

\bibitem{Gorkov16}
L.P. Gor'kov, V.Z. Kresin, Scientific Reports \textbf{6}, 25608 (2016).
\newblock \doi{10.1038/srep25608}

\bibitem{Duan14}
D.~Duan, Y.~Liu, F.~Tian, D.~Li, X.~Huang, Z.~Zhao, H.~Yu, B.~Liu, W.~Tian,
  T.~Cui, Scientific Reports \textbf{4}, 6968 (2018).
\newblock \doi{10.1038/srep06968}

\bibitem{Mahan90}
G.D. Mahan, \emph{Many-Particle Physics, second edition} (Plenum, New York,
  1990)

\bibitem{Alexandrov96}
A.S. Alexandrov, N.F. Mott, \emph{Polarons and Bipolarons} (WORLD SCIENTIFIC,
  1996).
\newblock \doi{10.1142/2784}

\bibitem{Friedel96}
{Friedel, Jacques}, J. Phys. I France \textbf{6}(12), 1493 (1996).
\newblock \doi{10.1051/jp1:1996170}

\bibitem{Schilling06}
J.S. Schilling, High Pressure Research \textbf{26}(3), 145 (2006).
\newblock \doi{10.1080/08957950600864401}

\bibitem{Raub84}
C.J. Raub, Platinum Metals Rev. \textbf{28}, 63 (1984)

\bibitem{Nucker87}
N.~N\"ucker, J.~Fink, B.~Renker, D.~Ewert, C.~Politis, P.J.W. Weijs, J.C.
  Fuggle, Z. Physik B -- Condensed Matter \textbf{67}, 9 (1987).
\newblock \doi{10.1007/BF01307300}

\bibitem{Tranquada87}
J.M. Tranquada, S.M. Heald, A.~Moodenbaugh, M.~Suenaga, Phys. Rev. B
  \textbf{35}, 7187 (1987).
\newblock \doi{10.1103/PhysRevB.35.7187}

\bibitem{Fujimori87}
A.~Fujimori, E.~Takayama-Muromachi, Y.~Uchida, B.~Okai, Phys. Rev. B
  \textbf{35}, 8814 (1987).
\newblock \doi{10.1103/PhysRevB.35.8814}

\bibitem{Barlingay90}
C.~Barlingay, V.~Garc\'{\i}a-V\'azquez, C.M. Falco, S.~Mazumdar, S.H. Risbud,
  Phys. Rev. B \textbf{41}, 4797 (1990).
\newblock \doi{10.1103/PhysRevB.41.4797}

\bibitem{Pelc15}
D.~Pelc, M.~Po\v{z}ek, V.~Despoja, D.K. Sunko, New Journal of Physics
  \textbf{17}(8), 083033 (2015).
\newblock \doi{10.1088/1367-2630/17/8/083033}

\bibitem{Little07}
W.~Little, M.~Holcomb, G.~Ghiringhelli, L.~Braicovich, C.~Dallera,
  A.~Piazzalunga, A.~Tagliaferri, N.~Brookes, Physica C: Superconductivity and
  its Applications \textbf{460-462}, 40  (2007).
\newblock \doi{10.1016/j.physc.2007.03.031}.
\newblock Proceedings of the 8th International Conference on Materials and
  Mechanisms of Superconductivity and High Temperature Superconductors

\bibitem{Bianconi87}
A.~Bianconi, A.~Castellano, M.D. Santis, P.~Rudolf, P.~Lagarde, A.~Flank,
  A.~Marcelli, Solid State Communications \textbf{63}(11), 1009  (1987).
\newblock \doi{10.1016/0038-1098(87)90650-8}

\bibitem{Valkov16}
V.V. Val'kov, D.M. Dzebisashvili, M.M. Korovushkin, A.F. Barabanov, JETP Lett.
  \textbf{103}, 385 (2016).
\newblock \doi{10.1134/S0021364016060114}

\bibitem{Si16}
Q.~Si, R.~Yu, E.~Abrahams, Nature Reviews Materials \textbf{1}, 16017 (2016).
\newblock \doi{10.1038/natrevmats.2016.17}

\bibitem{Garland69}
J.C. Garland, R.~Bowers, Phys. kondens. Materie \textbf{9}, 36 (1969)

\bibitem{Mukherjee15}
S.~Mukherjee, P.~Dasgupta, A.~Poddar, C.~Mazumdar, J Theor Appl Phys
  \textbf{10}, 75 (2015).
\newblock \doi{10.1007/s40094-015-0203-7}

\bibitem{Bussmann-Holder16}
A.~Bussmann-Holder, J.~K{\"o}hler, M.H. Whangbo, A.~Bianconi, A.~Simon, Novel
  Superconducting Materials \textbf{2}, 37 (2016).
\newblock \doi{10.1515/nsm-2016-0004}

\end{thebibliography}
\end{document}